\documentclass{aa}
\input epsf
\epsfxsize=\hsize

\begin{document}

\thesaurus{03 ( 02.02.1; 11.01.2; 11.05.1; 11.10.1; 11.14.1; 
13.18.1 )}
\title{Extragalactic radio sources with hybrid morphology: 
implications for
the Fanaroff--Riley dichotomy}

\author{Gopal-Krishna\inst{1}, Paul J. Wiita\inst{2} }

\offprints{Gopal-Krishna}

\institute{
National Centre for Radio Astrophysics, Tata Institute of Fundamental
Research, 
Pune University Campus, Post Bag No.\ 3, Ganeshkhind, Pune 411 007,
India 
(krishna@ncra.tifr.res.in)
\and
Department of Astrophysical Sciences, Princeton University,
Princeton, NJ 08544-1001, USA (wiita@astro.princeton.edu);\\
on leave from the Department of Physics \& Astronomy, Georgia State University,
 University Plaza,
Atlanta, GA 30303-3083, USA  (wiita@chara.gsu.edu)}

\date{Received 2 August 2000 / Accepted 15 September 2000}

\authorrunning{Gopal-Krishna \& Wiita}
\titlerunning{HYMORS and the FR dichotomy}
\maketitle

\begin{abstract}
We provide observational and theoretical perspectives on
the currently much debated issue of the Fanaroff-Riley (FR) morphological
dichotomy of extragalactic radio sources. In this context we introduce
a new, albeit rare, class of double radio sources in which the two lobes exhibit
clearly different FR morphologies.  It is argued that
such `HYbrid MOrphology Radio Sources', or HYMORS, could be 
used to effectively constrain the theoretical mechanisms
proposed for the FR dichotomy.
Basically, the existence of HYMORS supports explanations for the FR 
dichotomy based upon
jet interaction with the  medium external to the central engine, 
and appears quite difficult to
reconcile with the class of explanations that posit fundamental differences
in the central engine, such as black hole spin or jet composition, to be responsible 
for the two FR classes of double radio sources.

\end{abstract}

\section{Introduction}

A quarter of a century ago, using the first set of synthesis maps of 3CR radio
sources, Fanaroff \& Riley (\cite{fanaroff}) demonstrated that the morphology
of extended double radio sources undergoes a relatively sharp transition across
a critical radio luminosity, $L^*_R$, corresponding to $P_{{\rm 178} {\rm MHz}} \simeq
2 \times 10^{25}h_{50}^{-2}$ W Hz$^{-1}$ sr$^{-1}$.  
Most sources below this luminosity exhibit FR I
type structures, which are distinguished by diffuse radio lobes 
having their brightest regions within the inner half of the
radio source.
Such edge-dimmed radio sources include: fairly symmetrical 
twin-jets; fat doubles; 
Wide Angle Tail; and Narrow Angle Tail (or head--tail) sources.
On the other hand, the more powerful FR II type double sources show less
bending, and their  brightness peaks occur near the outer edges
of the two radio lobes, which are often identified as
hot-spots.  Interestingly, $L^*_R$
was found to lie near the observed break in the radio luminosity function 
of elliptical galaxies (Meier et al.\ \cite{meier79})
 and also to correspond to a transition in the properties of 
nuclear optical emission lines
(Hine \& Longair \cite{hine}).  
More detailed studies based on improved radio maps later
established that the  radio luminosity separating FR I from FR II sources
is actually a rising function
of the optical output of the parent elliptical galaxy, $L^*_R \propto
L_{\rm opt}^{1.7}$ (Owen \& White \cite{owen89}; Ledlow \& Owen \cite{ledlow}).

The origin of the FR I/FR II dichotomy continues to be a much debated 
outstanding issue in the astrophysics of extragalactic radio sources (e.g., Scheuer
\cite{scheuer96}).
Several authors have linked the morphological differences primarily to
the transition of an initially supersonic, but relatively weak, jet to a
transonic/subsonic flow decelerated substantially though entrainment of
thermal plasma within the inner ($\sim$ 1 kpc) region of the host
elliptical galaxy
(e.g., De Young \cite{deyoung}; Bicknell \cite{bicknell84}, \cite{bicknell94}, \cite{bicknell95}; Komissarov 1994; Bowman et al.\ 1996;
 Kaiser \& Alexander \cite{kaiser97}).
In contrast, several others have argued in favor of more fundamental
differences existing between the two classes, involving the nature of the
central engine (e.g., Baum et al.\ \cite{baum92};
Baum et al.\ \cite{baum95};   Reynolds et al.\ \cite{reynoldsa}; Meier et al.\ \cite{meier97},
Meier \cite{meier99}), or the possibility of composition of jet plasma being different,
with $e^-$--$e^+$ plasma inferred for FR I sources
(Reynolds et al.\ \cite{reynoldsb}), while $e^-$--$p$ jets may be preferred
for FR II sources (Celotti et al.\ \cite{celotti97}).
In our earlier work we had attributed the
FR I/FR II dichotomy primarily to differences in the jet's power/thrust
which, together with the properties of the circumgalactic medium, would
determine how soon the advance of the hot-spot 
becomes subsonic relative
to the ambient medium; this would lead to the disruption of the jet's collimation
due to its weakened Mach disk (Gopal-Krishna \& Wiita \cite{gkw88}; Gopal-Krishna
\cite{gk91}; also, Blandford \cite{blandford96}). Further evidence for this scheme,
which assumes no fundamental differences between either the central engines
or the jets of FR I and FR II sources, was presented
by Gopal-Krishna et al.\ (\cite{gk96}), based on a representative set
of {\sl Weak Headed Quasars}.  A growing body of VLBI observations
also indicates similar jet velocities in FR I and FR II
type sources close to the galactic nucleus (e.g., Giovannini et al.\ \cite{giovannini};
also Parma et al.\ \cite{parma}).

The goal of 
this communication is to present a new type of evidence in favor of
environmental factors being the primary determinant of the FR dichotomy.
In order to do so, it is necessary that we first present perspectives of the
observational status (Sect.\ 2) and theoretical inferences (Sect.\ 3) bearing on the question
of the FR I/FR II dichotomy.  We then
endeavor to constrain some of the proposed theoretical explanations
by introducing a new class of double radio sources, where the two lobes on the
opposite sides of the galactic nucleus exhibit clearly different FR morphologies
(Sect.\ 4).
Although rare, these {\sl HYbrid MOrphology Radio Sources}, or HYMORS, 
provide a valuable probe of the physical origin of the Fanaroff--Riley
morphological divide.  A brief report on this class was made in
Gopal-Krishna \& Wiita (\cite{gkw00}).  Our main conclusions are summarized
 in Sect.\ 5.

\section {Observational aspects of the FR I/FR II dichotomy}

Out of the vast amount of literature now available on this topic, we attempt
here to recapitulate some of the prominent distinctions claimed to
exist between the
two FR classes.  It is now well established that: FR II jets on kpc scales are 
distinctly more asymmetric and better collimated than FR I jets; the magnetic field 
in a FR II jet remains aligned
with the jet along most of its length, while in a FR I jet the magnetic
field is predominantly transverse on multi-kpc-scales
   (e.g.\ Bridle \& Perley \cite{bridle84}). Also, the radio 
nucleus is more prominent in the
FR I sources (e.g.\ Morganti et al.\ \cite{morganti}); however, the
 difference vanishes if FR I and FR II sources of the same radio luminosity are
considered (Zirbel \& Baum \cite{zirbel95}).  VLBI measurements of nuclear
jets, which often exhibit superluminal motions, strongly suggest
that FR II jets are relativistic on parsec scales, and there
are now many cases where FR I jets also appear to flow relativistically
on such scales (Giovannini et al.\ \cite{giovannini}; Bridle \cite{bridle96}; Laing 
\cite{laing99};
Biretta et al.\ \cite{biretta}; Xu et al.\ \cite{xu}).  
Doppler boosting can explain the facts that powerful FR II jets appear one-sided
while weaker FR I jets exhibit large brightness asymmetries only
near their origins,  and typically
have short, one-sided basal regions (Bridle \& Perley \cite{bridle84}; 
Parma et al.\ \cite{parma}).  
Moreover,  the Laing-Garrington (e.g., Garrington et al.\ \cite{garrington}) depolarization asymmetry 
is exhibited by the lobes of
some FR I sources as well as by the FR II sources in which it was discovered
(Parma et al.\ \cite{parma}; Laing et al.\ \cite{laing99}).
On the other hand, while the evidence
for FR II jets retaining their relativistic bulk velocities up to the multi-kpc
scale is very strong, with the brighter large scale jet always seen on
the same side as the nuclear jet and towards the less depolarized
radio lobe (Scheuer \cite{scheuer87}; Garrington et al.\ \cite{garrington}; Bridle
\cite{bridle96}), the diffuse 
nature of FR I sources, the brightness asymmetries of their jets decreasing with distance from 
the core, and often,
the strong bends seen in FR I jets, imply that much slower flows exist
on larger scales (e.g., O'Dea \cite{odea}; Feretti et al.\ \cite{feretti}; 
Laing et al.\ \cite{laing99}, and references therein).

Soon after their discovery it was noted that FR I
sources tend to be associated with dynamically evolved cD or D
type galaxies in clusters; in contrast, the hosts of FR II sources 
at similarly small redshift appear to avoid
rich clusters (Longair \& Seldner \cite{longair}; Seldner \& Peebles \cite{seldner};
Prestage \& Peacock \cite{prestage}; Owen \& Laing \cite{owen}; Zirbel \cite{zirbel97}), but more frequently have companion
galaxies and/or isophotal distortions which signify recent
galaxy mergers  (e.g., Heckman et al.\ \cite{heckman86}; Hutchings \cite{hutchings}; 
Baum et al.\ \cite{baum92};
Zirbel \cite{zirbel97}).  Furthermore, the 
hosts of FR I's are found to have an excess in optical size (relative to radio-quiet
ellipticals of the same optical magnitude) that correlates with $L_R$, while  
no such correlation is found for FR II's (Zirbel \cite{zirbel97}).  
Hill \& Lilly (\cite{hill}) argued that
the environment of FR II sources changes with cosmic epoch, in the sense that
by moderately high redshifts ($z \sim 0.5$), they begin to be found inside 
rich clusters; also see Zirbel (\cite{zirbel97}).  Recent
work (McLure \& Dunlop \cite{mclure}; Wold et al.\ \cite{wold}), however, indicates that there may
be negligible cosmological evolution of the environment, and that
powerful AGN do not really avoid clusters, even at small redshifts.  Similarly, the
well known
statistical trend for the host galaxies of FR II sources to be about 0.5 magnitude
fainter than those of FR I sources (Lilly \& Prestage \cite{lilly}; 
Prestage \& Peacock \cite{prestage};
Smith \& Heckman \cite{smith}; Owen \& Laing \cite{owen}) has recently been explained
as a selection effect arising from a combination of the $\sim L_{\rm opt}^2$ dependence of
the radio power at the FR I/FR II transition and the steepness of the radio
luminosity function of elliptical galaxies (Scarpa \& Urry \cite{scarpa}).  

Additional differences between the optical properties of the host galaxies 
of FR I and FR II sources have been noted.
Although FR II's exhibit roughly an order
of magnitude stronger optical line emission than do FR I's of the 
same radio luminosity, the
optical line emission seems to correlate with the host's optical magnitude
only for FR I's (Baum et al.\ \cite{baum92}; Zirbel \& Baum \cite{zirbel95}).  
The indicated internal origin of the gas in FR I's would be consistent with the
recently inferred origin of dusty material detected in FR I sources,
which appears to be generated either within the elliptical host itself,
or acquired in close encounters with gas-rich galaxies (in contrast to
an origin through violent mergers in the case of FR II's) (de Koff et al.\
\cite{dekoff}).  Further, in contrast to FR II's,
only a weak correlation of optical line emission with $L_R$ is 
found for FR I's (Baum et al.\ \cite{baum95}).  Spectroscopic observations have 
indicated that the kinematics of the ionized gas in FR I's is turbulence dominated, 
while some ordered bulk rotation is
present in the ionized gas associated with FR II hosts, and this rotation axis tends
to be aligned with the radio axis (Baum et al.\ \cite{baum92}). The emission line
ratios of these ``rotator''-type nebulae found in FR II sources are
consistent with photoionization by the nuclear continuum, since 
 the [O I] 6300, [N II] 6584 and [S II] 6717 forbidden lines
are very weak relative to H$\alpha$.
Recent HST images of FR I nuclei reveal
a deficiency of the thermal UV emission which is usually attributed to
the nuclear accretion disk; this may
account for the faintness of nuclear optical line emission (Chiaberge et al.\
\cite{chiaberge99};
also Zirbel \& Baum \cite{zirbel95}).
Interestingly, Chiaberge et al.\ (\cite{chiaberge00}) also found a similar situation
for some of the  FR II
sources of modest radio luminosity.

Another notable difference between FR I and FR II hosts
pertains to the amount of mid/far-IR (MFIR)
emission: for samples matched in radio luminosity, FR II hosts 
typically have $\sim 4$ times stronger MFIR emission, perhaps
attributable to nuclear starbursts induced by galaxy mergers,
which is also consistent with the higher rate of occurrence of optical distortion
found for the FR II host galaxies
(Heckman et al.\ \cite{heckman94}).

\section{Some theoretical considerations related to the morphological dichotomy}

\subsection{Jet deceleration}

In light of the evidence for relativistic jet velocities persisting up
to multi-kpc scales in 
FR II sources, coupled with the likelihood that the jets in both FR types start
out with bulk relativistic speeds, many theoretical studies have stressed the
need for deceleration of the jet flow in FR I sources.  Begelman (\cite{begelman}) argued that
viscous dissipation in jets can balance adiabatic heating and cause a
rather rapid deceleration of weaker jets to transonic or subsonic speeds.
These jets could remain undisrupted for substantial distances, thereby yielding
typical FR I morphologies, provided the external pressure gradient is appropriately
steep. 
De Young (\cite{deyoung}) noted that the Owen-Ledlow transition from FR II to FR I at a
fixed $L_R$ and increasing $L_{\rm opt}$ could correspond to a supersonic (perhaps
relativistic) jet being severely decelerated in the inner $\sim 1$ kpc of the parent
galaxy, where more gas is likely to be available for entrainment.  Plausibly, enough
of such gas
could arise from stellar winds, or perhaps from
cooling flows onto the cD galaxies which often host FR I sources.

Bicknell (\cite{bicknell84}, \cite{bicknell94}, \cite{bicknell95}) focused on 
the idea that turbulent entrainment of cool 
interstellar medium at the jet boundary could dramatically decelerate
a jet.  His original work (Bicknell \cite{bicknell84}) assumed non-relativistic FR I
jets throughout, and ran into some difficulties (e.g., Laing et al.\ \cite{laing99}).
But the later model (Bicknell \cite{bicknell94}, \cite{bicknell95}), which assumed initially relativistic 
jets which eventually come into pressure balance with the external medium, is quite
successful in reproducing many aspects of the observations.  Bicknell
 argued that the instability to
Kelvin-Helmholtz modes that would produce jet flaring, and thus an FR I morphology,
tended to occur at Mach numbers of $\sim 2$ or flow velocities of
$\sim 0.6$c.  This result was shown to hold for wide ranges of initial relativistic 
velocities and of
initial ratios of cold to relativistic matter in the jet (Bicknell \cite{bicknell95}).
By incorporating the known empirical relationships between the optical
and  X-ray properties of elliptical galaxies, Bicknell's (\cite{bicknell95}) 
model could effectively account for the observed slope (and approximate 
intercept) of the
Owen-Ledlow boundary for the FR I/FR II transition in the $L_R$--$L_{\rm opt}$
plane.
Self-similar models for radio source growth by Kaiser \& Alexander (\cite{kaiser97})
feature a turbulent shear layer that could disrupt weaker jets,
turning them from FR II into FR I type morphologies if the external
density gradient was rather shallow.
Komissarov (\cite{komissarov94}) and Bowman et al.\ (\cite{bowman}) also considered entrainment
as leading to jet deceleration in FR I sources.  Bowman et al.\ (\cite{bowman})
argued that FR I plasma was initially hotter and stressed the importance of cool
stellar matter directly swept up by the jets.  They showed that this could
produce substantial deceleration, even if the jets were highly relativistic initially,
without causing a precipitous dissipation of the jet's kinematic power
and the ensuing dramatic brightening, which is not observed (see, Scheuer
\cite{scheuer83}).

Observational support for the models invoking decelerating
FR I jets comes from the anti-correlation found between the apparent brightness
ratio and the width ratio of the twin jets in FR I sources, which is readily
understood in terms of Doppler boosting of a centrally peaked
velocity profile (Laing et al.\ \cite{laing99}, and references therein).  
A wide variety of the observed characteristics of the radio jets
in FR I sources, such as the emission gaps seen near the nucleus (Komissarov
\cite{komissarov90}),
the asymmetries in apparent emission from the two jets, and their 
magnetic field patterns, are reasonably explained if the
jets in these sources consist of
a narrow ``spine'' of relativistic flow
with a predominantly transverse magnetic field, surrounded by a slower
moving ``sheath'' contaminated by entrained material (a shear layer) 
where the magnetic field is stretched into a predominantly longitudinal 
configuration (Laing \cite{laing93}, \cite{laing96}; Laing et al.\ \cite{laing99}). 
 This picture is in accord with
Bicknell's (\cite{bicknell95}) transonic relativistic jet models which are confined
by external pressure at large distances, and where the deceleration
usually occurs within $\sim 2$ kpc of the core.

\subsection{Jet composition}

Total energy and synchrotron radiation constraints led Celotti \& Fabian
(\cite{celotti93}) to conclude that FR II jets were made of $e^-$--$p$ plasma,
since they argued that $e^-$--$e^+$ plasma of the required density
would yield too much annihilation radiation.  On the other hand,
Reynolds et al.\ (\cite{reynoldsb}) used similar energetic and radiation constraints
to conclude that the jet in the FR I source M87 was likely to be
made of $e^-$--$e^+$ plasma.  A similar argument favors an electron--positron
jet in the Optically Violently Variable Quasar 3C 279 (Hirotani et al.\ 
\cite{hirotani}).  If all of these arguments
are taken at face value, one might infer that the main difference between
FR I and FR II sources lies in the composition of the jet plasma,
and this would imply the existence of a qualitative difference
between their central engines.  However, evidence for the presence of
pair plasma jets, even in FR II sources, comes from the interpretation of
the radio power--linear-size (P--D) diagram in terms of a model for 
self-similar growth 
of double radio sources
(Kaiser et al.\ \cite{kaiseretal97}).
Furthermore, there are viable alternatives to the annihilation
constraint invoked by Celotti \& Fabian (\cite{celotti93}) to argue against pair
plasmas in FR II jets; for example, the earliest stage of the energy 
transport could be predominantly via Poynting flux
(Reynolds 1999, private communication), in which case the radiating
relativistic matter in all jets could indeed be essentially an
 $e^-$--$e^+$ plasma.

The ``spine/sheath'' model (Sect.\ 3.1) is broadly reminiscent of the two-fluid-type 
configuration for
jets, put forward by Pelletier \& Roland (\cite{pelletier}). (See also Sol et al.\
\cite{sol}.) 
 They suggest that the spine of the jet
is relativistic, at least on parsec scales, and is composed of a pair
plasma, while the outer sheath is made of $e^-$--$p$ plasma, and
carries the bulk of the energy to the outer lobes.

\subsection{Galactic mergers}

Substantial isophotal distortions are observed in the
ellipticals hosting both FR I and FR II sources, strongly
implying that galactic encounters/mergers have occurred (Heckman et al.\ \cite{heckman86};
Colina \& de Juan \cite{colina}).   However, the 
distinctive sharpness of the distortions seen in 
the FR II hosts (Smith \& Heckman \cite{smith}), combined
with the presence of strong optical emission lines  and
significantly higher MFIR emission (Heckman et al.\ \cite{heckman94})
suggests the occurrence of a starburst due to merger of  a gas rich spiral
with the elliptical host (Smith \& Heckman
\cite{smith}; Colina \& de Juan \cite{colina}).
In contrast, elliptical--elliptical mergers have been
invoked in the case of FR I galaxies (Colina \& de Juan \cite{colina}). 

  Consecutive mergers of galaxies containing central supermassive
black holes (SMBHs)
could produce multiple SMBH systems.  Such triple systems usually
become unstable and eject a single black hole in one direction, and
the recoil sends the surviving binary black hole system in the opposite
direction; this is the core idea of the
gravitational slingshot model for radio source production
(e.g., Saslaw et al.\ \cite{saslaw}). In this scenario, FR I
sources correspond to SMBHs ejected at less than the escape velocity
from the merged host galaxy, while FR II sources arise from
SMBHs that do escape (Valtonen \& Hein\"am\"aki \cite{valtonen}).
Since this picture naturally produces different velocities for
SMBHs of different masses, it could both produce HYMORS,
and even make actual predictions as to their frequency.
Valtonen \& Hein\"am\"aki (\cite{valtonen}) also argue that the
slingshot model can roughly account for the dependence of
$L^*_R$ on $L_{\rm opt}$, as well as for many other properties of 
radio galaxies.

\subsection{Spin of the central engine}

Some hydromagnetic process is now widely believed to
be responsible for launching relativistic jets from the vicinity of
the accretion disk/supermassive black hole combination which is believed
to constitute the central engine in all AGN, although the
details remain highly contentious (e.g., Scheuer \cite{scheuer96}; Wiita \cite{wiita}).
A possible hint that the angular momentum of the central
engine is important in launching powerful FR II jets
comes from the rotational kinematics of the (presumably accreted)
ionized gas
observed in FR II host galaxies (Baum et al.\ \cite{baum92}).
The idea that a merger of two SMBHs belonging to a merged 
pair of elliptical galaxies could yield a single rapidly spinning SMBH, which 
propels powerful relativistic jets, was advocated by Wilson \& Colbert (\cite{wilson}).
While the black hole spin may well be an important ingredient for 
ejection of powerful
jets, the existence of HYMORS (Sect.\ 4) disfavors differences in the SMBH spin
 as the principle mechanism for the FR dichotomy.
 
One basic class of scenarios involves variants of the Blandford-Znajek (\cite{bz}, B-Z)
mechanism, which could tap the SMBH's rotational energy via
magnetic field lines threading the SMBH horizon.  While extremely
efficient in principle, and capable of providing powerful radio jets with
minimal optical thermal emission if an ion-supported torus forms in the innermost region
(Rees et al.\ \cite{rees}), the viability of this mechanism has recently
been questioned on several grounds.  Ghosh \& Abramowicz (\cite{ghosh}) argued
that the strength of the magnetic field that could actually thread
the SMBH horizon may have been substantially overestimated.  Even if the B-Z mechanism
works, Livio et al.\ (\cite{livio}) have claimed that the emitted
power is dominated by energy output from the inner disk regions, at least
for the standard thin accretion disks, and therefore the efficiency is
much reduced.  This last limitation may be overcome if the accretion
disk is actually thick in the inner regions (Armitage \& Natarajan \cite{armitage}).
An additional potential problem for the B-Z mechanism was 
recently pointed out by Li (\cite{li}); he argued that the plasma screw instability 
must set in and this implies that even if the B-Z mechanism does work locally, any jet it launches
would be severely limited in its overall length.

While none of the above critical arguments can be considered to be
watertight, they tend to support the alternative basic
scenario, which involves hydromagnetic launching of the jets
from the accretion disk, rather than from the immediate vicinity of the
SMBH (e.g., Blandford \cite{blandford94}).  Most of 
these disk-origin models of jets (e.g.\ Appl \& Camenzind \cite{appl};
Chiueh et al.\ \cite{chiueh}) can be considered
to be variants of the Blandford-Payne (\cite{bp}) scheme.
However, it should be noted that if the screw-instability argument of Li (\cite{li}) turns
out to be valid, it probably also applies to disk-launched
jets and would cause difficulties for any MHD dominated jet formation process.

One possible approach  for producing asymmetric jets  by a single
central engine was proposed by Wang et al.\ (\cite{wang}).  They took
a semi-analytical approach to the force-free Grad-Shafronov equation
and found solutions in which the bulk of the power was carried by
the Poynting flux, while most of the angular momentum in the jet
was carried by the magnetic fields emerging from the accretion disk.
Wang et al.\ (\cite{wang}) found that substantially more thrust could flow
off of one side of the disk than the other if sufficiently large
asymmetries could be maintained in the magnetic field within the disk.

The idea that the accretion disk corona can generate two fundamentally
different types of jet has been proposed recently (Meier et al.\ \cite{meier97};
Meier \cite{meier99}).  Fast (highly relativistic, FR II) jets are ejected
when the coronal plasma is unbound by the magnetic field, while
slower (transrelativistic, FR I) jets, moving at roughly the
disk's escape velocity, are produced when the corona is inertially bound
to the SMBH.
The original version of this ``magnetic switch'' mechanism (Meier et al.\ \cite{meier97})
could explain how jets of very different speeds arise from otherwise
similar sources, but this version fails to explain how these
differences can persist over the extended lifetimes of FR I sources
(Meier \cite{meier99}).  In his revised scenario, Meier (\cite{meier99}) argued that the
difference in radio jet power among galaxies of the same mass arises from
different speeds of rotation of the magnetic field lines associated
with their central engines,
which are in turn produced by different spin rates of their SMBHs.
(The idea that the SMBH spin was critical had been put forward already
by Baum et al.\ (\cite{baum92}, \cite{baum95}) based 
on observational inferences about the merger of the host galaxy with a
high angular momentum, gas-rich, disk galaxy in the case of FR II
sources.)
The transition occurs at a critical power when the MHD luminosity,
$L_{\rm MHD} = B_{p0}^2 R_0^3 \Omega/2$, (where $B_{p0}$ is the poloidal
magnetic field, $R_0$ is the size of the magnetic rotator, and $\Omega$
its angular velocity)
exceeds a critical luminosity, defined as the liberation of an
escape energy in a free-fall time:
\begin{equation}
L_{\rm crit} = {\frac{E_{\rm esc0}} {\tau_{\rm ff0}}} = 4\pi \rho_{c0} R_0^2
\Bigl({\frac{GM}{R_0}}\Bigr)^{3/2}.
\end{equation}
In Meier's scenario,
this magnetic switch luminosity
plays the same role in MHD acceleration as does the Eddington luminosity
in radiative acceleration.  
Note that this magnetic switch model relies on extracting substantial power
from a portion of the accretion disk extending within the ergosphere,
thereby avoiding the problem highlighted by Livio et al.\ (\cite{livio}).  
This model is capable of yielding a decent match to the
Owen-Ledlow variation in the value of $L^*_R$ with $L_{\rm opt}$
in terms of a critical SMBH spin (Meier \cite{meier99}). 

  The possibility that Advection Dominated Accretion Flows (ADAFs;
e.g., Narayan \& Yi \cite{narayan}), which are inefficient radiators, 
are present in FR I 
radio galaxies was first proposed by Reynolds et al.\ (\cite{reynoldsa}).
In this picture, which is an intriguing option (e.g., Jackson
\cite{jacksonc}),
standard thin accretion
disks, which radiate more efficiently, yield FR II radio sources
(Reynolds et al.\ \cite{reynoldsa}).  A good fit
to the low-frequency radio and X-ray emission of M87 (Reynolds
et al.\ \cite{reynoldsa}) as well as to that of several quiescent ellipticals
(Di Matteo \& Fabian \cite{dimatteo97}; Mahadevan \cite{mahadevan})
could be attained using ADAF models.  However, the ADAF models grossly overpredict
the high-frequency radio and sub-mm emission from these quiescent
galaxies unless: the magnetic fields are very much below equipartition;
or there is enough cold material for free-free absorption of the 
synchrotron absorption
to be very important; or powerful winds remove much of the energy,
angular momentum and mass from the inner part of the accretion flow
(Di Matteo et al.\ \cite{dimatteo99}).

\section{HYMORS: a new observational clue to the FR dichotomy}

As mentioned in Sect.\ 1, the remarkable differences between a
wide range of characteristics of the FR I and FR II sources, as summarized
in Sect.\ 2, have led several authors to the  viewpoint which ties the origin
of these differences to the properties of the central engine itself 
(Sects.\ 3.2--3.4). It is clearly important to confront
this somewhat radical stand with any discriminating observational 
results available.
One possible strategy is to look for double radio sources whose radio
structures on the two sides of the nucleus exhibit {\it different}
Fanaroff-Riley morphologies.  Even a few examples of such clearly hybrid
morphology double radio sources would call into question models
that attribute the FR dichotomy to the properties of the central engine,
since in a given double source both radio lobes are presumably caused by
a single central engine. On the other hand, such a hybrid morphology may
be readily accommodated within a scenario where the ambient media on the
two sides of the nucleus have sufficiently dissimilar properties so as to
impose different fates upon the two jets emanating from the nucleus.

Following the above reasoning, we have carried out a substantial,
though certainly not exhaustive, search
of the published literature and have located several cases of {\it  HYMORS},
which  illustrate our point. Below we briefly comment on these individual
examples, whose basic properties are summarized in Table 1.

\noindent{\bf 0131$-$367} (NGC 612, $z$ = 0.029):  This bright SO galaxy
with a prominent dust-lane is the host of a prominent double radio source (Ekers et
al.\ \cite{ekers}). The hybrid nature of its radio structure is evident
from its 5 GHz VLA map which shows a weak core flanked by two radio
lobes; the eastern lobe has a bright hotspot near its outer edge
(FR II type), whereas the western lobe exhibits a  jet-like
structure which widens steadily and fades into a diffuse radio plume (FR I)
(Fig.\ 1; Morganti et al.\ \cite{morganti}).

\noindent{\bf 0521$-$364} ($z$ = 0.055): This well-known blazar, found
to be a source of $\gamma-$rays above 100 GeV (Thompson et al.\ \cite{thompson}),
is another fine example of hybrid radio morphology. It consists
of a radio/optical synchrotron jet which does not terminate in a hot spot,
and a bright radio hot spot on the counter-jet (SE)  side, all embedded
in a radio halo (Fig.\ 2; Keel \cite{keel}).

\noindent{\bf 1004+130} (4C+13.41, $z$ = 0.240): The hybrid morphological
nature of
this quasar is apparent from its 5 GHz VLA map made by Fomalont (\cite{fomalont};
Fig.\ 3). The radio lobe westward of the bright nuclear core is strongly
edge-brightened, typical of FR II sources. In contrast, the eastern lobe
is clearly edge-darkened (FR I type), and its structure is dominated by a jet
which progressively fades away from the nucleus.

\noindent{\bf 1452$-$517} ($z$ = 0.016): A 843 MHz MOST map of this giant
radio galaxy was made by Jones (\cite{jones86});
 also see Jones \& McAdam (\cite{jones92}). Whilst the N lobe is
edge-brightened (FR II type), the other lobe is edge-darkened (FR I)
(Jones \cite{jones86}). A higher resolution map is needed to ascertain if the
elongated radio feature seen in the S lobe is indeed a
jet. An alternative possibility in the context of such giants is that
the emission peak recessed from the outer edge could be the current
working surface of a rejuvenated jet which had already reached a much
greater extent but was cut off by the instabilities which are particularly
likely to
afflict such giant, old, radio sources (e.g., Hooda et al.\ \cite{hooda}).

\noindent{\bf 1726$-$038} (4C$-$03.64): This double source is a
good example of a HYMORS. As seen from a recent 4.9 GHz VLA map
(Fig.\ 4; Jackson et al.\ \cite{jackson}), the extremity of the NE lobe
is marked by a bright hot spot (FR II type). In contrast, the SW side of
the nucleus exhibits a prominent jet which progressively fades away
outwards, a standard FR I pattern.  Unfortunately,
the authors provide no redshift measurement; using the digitized POSS
we have attempted to obtain a crude estimate of the redshift. We
tentatively identify the source with an elliptical galaxy of
approximately 10 arcsec extent.
This angular size of the host galaxy suggests a redshift of $\sim 0.05$,
taking
the intrinsic diameter to be 10 kpc and a Hubble
constant of $H_0$ = 75 km s$^{-1}$ Mpc$^{-1}$.

\noindent{\bf 2007$+$777} ($z$ = 0.342): This moderately distant, core
dominated BL Lacertae object has been mapped with the VLA at 1.5 GHz
by Murphy et al.\ (\cite{murphy}). In this source the eastern side
of the nucleus exhibits a prominent hot spot, whereas the western side is
marked by a jet which gradually fades into oblivion at about 10 arcsec
from the nucleus.

Some high redshift examples of  HYMORS can be identified from 
the quasar sample imaged by Lonsdale et al.\ (\cite{lonsdale}), 
using the VLA at 5 and 15 GHz.
{\bf 0038$-$019} (4C $-$02.04; $z$ = 1.690) is a 
good
example of hybrid morphology, with a hot-spot in the N lobe and a southward jet
without a terminal hot-spot.  {\bf 1258$+$404} (3C 280.1, $z$ = 1.659) 
has a hot-spot on the NW and only a jet to the SE.
{\bf 1323$+$655} (4C 65.15, $z$ = 1.618) is another possible HYMORS, 
with a compact hot-spot to the NE and a bent jet on the SW.  The peculiar
structures of
the first and third of these sources were noted by Lonsdale et al.\ (\cite{lonsdale}).

Among the AGN associated with disk galaxies, a HYMORS-like
source has recently been mapped in the Seyfert galaxy, Mrk 3 (UGC 3426).
Mrk 3 has an extremely low level of nuclear radio activity and
its overall radio size is only $\sim$ 0.5 kpc (Kukula et al.\ \cite{kukula}).
The E jet shows knots close to the nucleus and fades outwards in
an FR I pattern, while the W jet terminates in a typical FR II hot spot
and lobe.  

In addition, 0905$-$353, which was mapped at low-resolution
by Jones \& McAdam (\cite{jones92}), is another possible HYMORS. We plan
to employ the high
resolution and sensitivity of the the Giant Metrewave
Radio Telescope to ascertain if this object
 indeed has a hybrid morphology.

  Although Fanaroff \& Riley (\cite{fanaroff}) classified radio sources solely in terms
of whether the separation of points of peak intensity were less than (FR I)
or greater than (FR II) half of the largest size of the source, more
subtle classification schemes have since been proposed.  Probably the most widely
adopted of these separates double sources into categories depending upon: (a) whether
the extended emission is best described as plume-like or bridge-like; and
(b) whether
the lobes possess compact features or not, and if they do, whether the compact emission
is dominated by hotspots, weak jets or strong jets (Leahy \cite{leahy93},
\cite{leahy00}; Laing \cite{laing93}).  Since almost all FR II sources would
fall into a
the category which is dominated by bridges on large scales and hotspots on small scales, this
scheme mainly serves to stress the wide range of types commonly clubbed together in
the FR I fold, a few of which might be called `FR 1.5' 
(Leahy \cite{leahy00}).

Other types of  structures intermediate between FR I and FR II morphologies
have been noted earlier (Zirbel \& Baum \cite{zirbel95} and references therein). 
  Owen \& Laing (\cite{owen89}) discuss a small group of ``Fat Doubles'',
with bright outer rings and rounded lobes, which they argue are best considered
as FR I/II transitional sources.  We would not consider any of these as an
example of a HYMORS.
In their list of some 334 sources, for which they had good enough maps to
classify 212, Zirbel \& Baum (\cite{zirbel95}) list only 7 sources as being
`FR I/II' (in their paper this means different categorizations had 
appeared in the literature)
and only 3 as being `Transitional'.  
Published maps could be located for 7 of these 10;
we would call four of these clearly FR II, one clearly FR I, 
one (3C 17) as being very confused,
perhaps involving a chance superposition of an FR II and an FR I (Morganti
et al.\ \cite{morganti}), and
only one case (3C 15) that might legitimately be transitional; however, it did not
fit our definition of a HYMORS.  Nor does PKS 1313$-$33, a source Morganti et al.
(\cite{morganti}) call transitional.
In addition, transitional type sources housing plumes and tails 
along with well collimated jets and weak hot spots, have been seen in three
intermediate power radio galaxies (Capetti et al.\ \cite{capetti}) and in some low luminosity
radio sources (Parma et al.\ \cite{parma}); again, none of these are good examples
of HYMORS.  Thus, while the sobriquets `FR 1.5' and `FR I/II' have appeared
occasionally in the literature, their meanings are imprecise, and we consider ourselves
justified in introducing the more specific term HYMORS to describe sources with 
definite
FR I morphology on one side and distinct FR II morphology 
on the other.

\section {Concluding remarks}

Our main objective in this study was to highlight the
existence of double radio sources whose structures are characterized by
a hybrid morphology in terms of the Fanaroff-Riley classification scheme.
Table 1
summarizes the basic radio and optical data for six good examples
of such HYMORS (Sect.\ 4).
It is seen that the hybrid radio morphology can be associated with all 
three major classes of radio powerful AGNs (i.e., galaxies, quasars and 
 BL Lac objects), and with radio sources extending from galactic up to
megaparsec dimensions. Although the radio powers of these HYMORS 
are below or around the critical value for the FR I/FR II transition, this 
probably is due to the modest redshifts of these sources; their counterparts 
may well exist at high redshifts ($z > 1$, Sect.\ 4). From Figs.\ 1--4 it may also 
be noted that in some of the cases the FR I classification of a radio lobe 
is based on the detection of a radio jet without a terminal hot spot; in the 
other cases a diffuse radio lobe without a hotspot (FR I type) is 
found associated with a jet.    We note that since the missing hotspot 
in HYMORS is often on the side with the jet, the morphological
asymmetry cannot be explained by postulating even an implausibly
strong Doppler boosting of the hotspot radiation.

  Because we have examined very heterogeneous samples of radio maps
in order to produce our set of HYMORS, it is not possible to
give an accurate estimate of their frequency.  We examined roughly 1000
radio maps overall to come up with the 6 good examples highlighted above.
Among relatively well defined samples, we found:  1 HYMORS out of 181 3CRR 
galaxies considered by Laing et al.\ (\cite{laingetal}); 1 case among the
150  4C maps presented by
Jackson et al.\ \cite{jackson}; and 0 cases in the 98 maps of steep
spectrum 4C sources considered by Rhee et al.\ (\cite{rhee}).
These are consistent with very low rates of less than about $1\%$.
The higher redshift sample of Lonsdale et al.\ (\cite{lonsdale}),
however, yielded 3 HYMORS out of 70 sources.

Although rare,  HYMORS can serve as a very 
useful discriminator between the wide range of theories that have been 
put forward to explain the origin of the FR dichotomy.  The mere existence of 
HYMORS seems to disfavor the class of models that posit fundamental differences 
between the central engines, such as spin or jet composition, as being 
the dominant cause for the two morphological 
types (Sects.\ 1, 3).  The scheme of Wang et al.\ (\cite{wang}) (Sect.\ 3.4), where the
central engine is argued to be capable of ejecting jets of
grossly unequal power, could conceivably be reconciled
with the existence of HYMORS.   However, even if such 
     an asymmetry could somehow be maintained over most of the source lifetime,
there is no significant evidence in 
our sample of HYMORS
for the positively correlated asymmetries in radio luminosities and lobe
lengths that this scenario would predict.   (If one leaves the standard
jet paradigm and considers gravitational slingshot scenarios (e.g.,
Valtonen \& Hein\"am\"aki \cite{valtonen}), then some HYMORS would
be expected, though the predicted lobe length asymmetry cannot be verified using
     the present small sample.)

Thus, at least in the case of HYMORS, it appears that some type of jet-medium 
interaction on kiloparsec scales is playing a crucial role in creating the
morphological asymmetry about the nucleus.     If the  asymmetric
jet/medium interaction is such that the jet
collimation is quite different on the
two sides, the side with poorer collimation, and therefore more rapid 
slowdown, would lose its hotspot sooner. Earlier, environmental
asymmetries have been argued to be important for the Compact
Steep Spectrum radio sources (Gopal-Krishna \& Wiita \cite{gkw91}; Saikia et al.
\cite{saikia}; Jeyakumar et al.\ \cite{jeyakumar}).  A more 
extensive search for HYMORS, allowing a reliable statistical estimate of 
their frequency and structural asymmetries, is  likely to provide
a deeper insight into the origin of the 
Fanaroff-Riley morphological dichotomy. 

  A recent study appears to  underscore
      the dominant role of accretion in the jet formation.
Serjeant et al.\ (\cite{serjeant}) show that for steep spectrum 
quasar cores, radio and optical outputs are strongly correlated; this implies a close
link between the formation of jets and accretion onto the SMBH,
improving on the similar argument made by Rawlings \& Saunders (\cite{rawlings};
see also Falcke \& Biermann \cite{falcke}).

Finally, in the context of HYMORS it may be pertinent to quote
Lilly \& Prestage (\cite{lilly}): ``It is important to stress that the environment
must be influencing not just the outer radio lobes, but also nuclear
phenomena, such as the strength of the optical emission lines and the
properties of the radio jets.''  We surmise that the observed differences 
between the properties
of the host galaxies of FR I and FR II sources  engender the
dichotomy by creating different environmental conditions to be encountered
by the jets, rather than by producing fundamentally different kinds of
central engines.

\begin{acknowledgements}
We thank Drs.\ E.\ Fomalont, N.\ Jackson, P.\ Jones, W.\ Keel, R.\ Morganti, and 
D.\ Murphy
and their publishers for permission to reproduce
their radio maps and for providing electronic versions of the same,
when available. 
PJW appreciates the hospitality at the National Centre for Radio Astrophysics
 where this work was begun and
G-K  gratefully acknowledges travel support from Georgia State University and the
hospitality at Princeton University where this paper was written.
  This research has made use of the NASA/IPAC
Extragalactic Database (NED) which is operated by the Jet Propulsion Laboratory,
California Institute of Technology, under contract with the National Aeronautics and
Space Administration. 
This work was supported in part by NASA grant NAG 5-3098 and Research
Program Enhancement funds at GSU.
\end{acknowledgements}

\newpage

\begin{table}
\caption{Properties of the HYMORS}
\begin{tabular}{ccccc}

%\tableline
%\tableline
 
Object &  $z$ & Size & Size$^a$ & Log (L$_R$) \\
 & &(arcmin)&(kpc) &(1 GHz) W/Hz$^a$ \\
%\tableline

0131$-$367 & 0.029 &   14.2 & 483 & 25.4 \\
0521$-$364 & 0.055 &  0.3   & 22  & 25.4\\
1004$+$130 & 0.240 & 1.8    & 524 & 26.3\\
1452$-$517 &  0.08 & 20.3   & 812 & 25.4\\
1726$-$038 &  (0.05)$^b$&0.6    & 35  & 24.8\\
2007$+$777 & 0.342 & 0.5    &  213 &  24.8 \\

%\tableline
\end{tabular}

$^a$$H_0$ = $75$ km s$^{-1}$ Mpc$^{-1}$,  
$q_0 = 0.5$, spectral index = $-1$.

$^b$Estimated redshift (Sect.\ 4).
%\end{tabular}
\end{table}

\clearpage

\begin{figure}
\epsffile{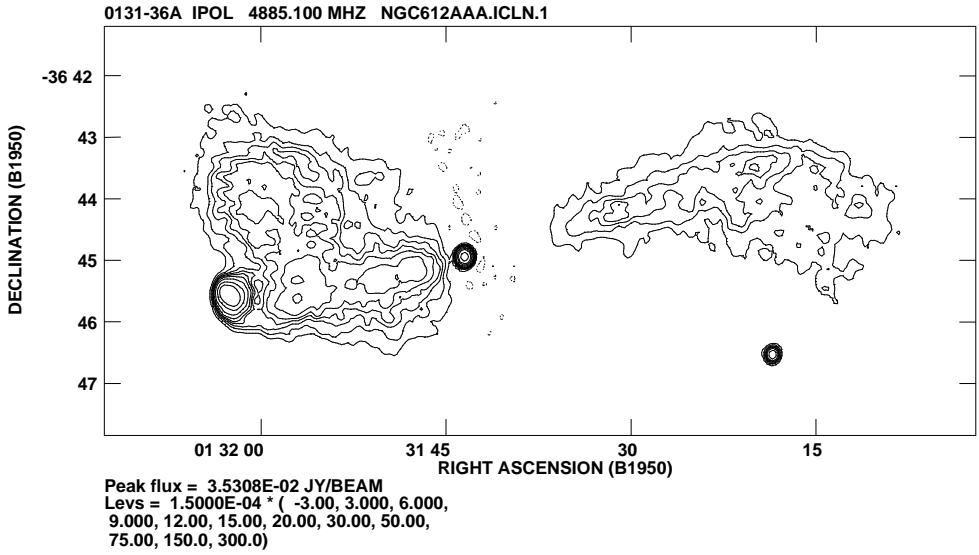}
\caption{Maps reproduced from the literature showing the hybrid morphology
of some double radio sources:  0131$-$367, reprinted with permission from
Morganti et al.\ (1993), copyright, Royal Astronomical Society.}
\end{figure}
%\clearpage

\newpage
\begin{figure}
\epsffile{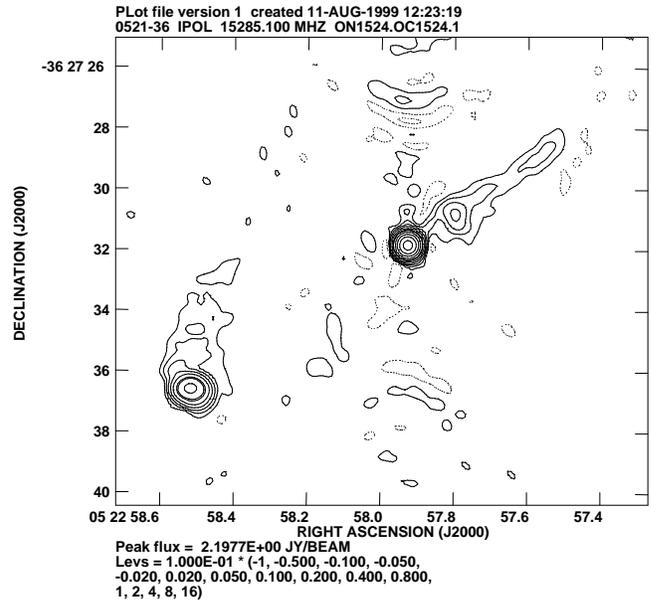}
\caption{ 0521$-$364,
 reprinted with permission from Keel (1996), copyright, American Astronomical
Society.}
\end{figure}
\clearpage
%\newpage
\begin{figure}
\epsffile{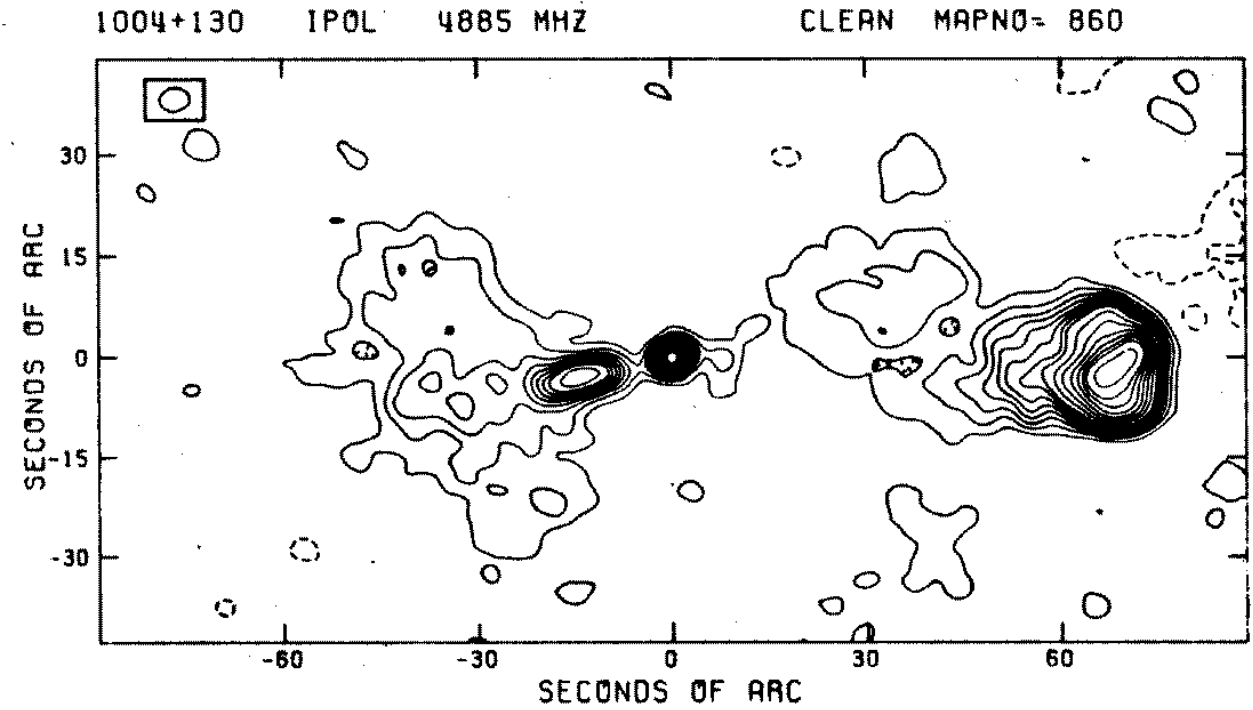}
\caption{  1004$+$130,  reprinted with permission from Fomalont (1982),
copyright, Kluwer Academic Publishing.}
\end{figure}
%\clearpage
\newpage

\begin{figure}
\epsffile{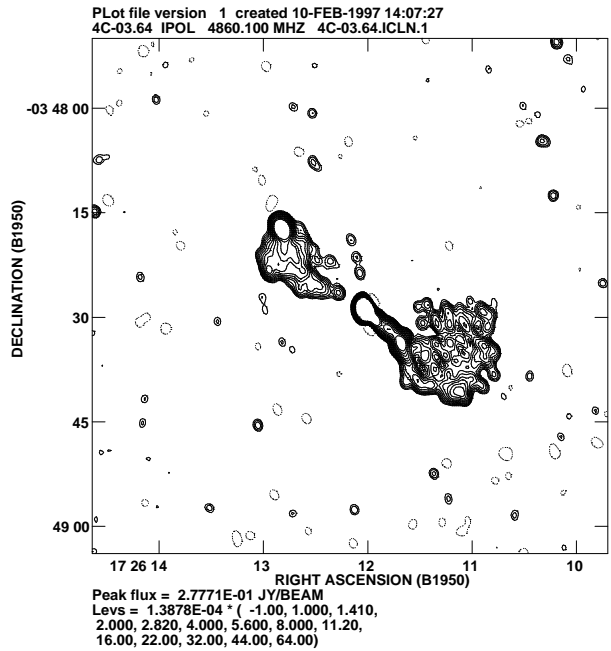}
\caption{1726$-$038,
reprinted with permission from Jackson et al.\ (1999), copyright, European 
Southern Observatory.}
\end{figure}
\clearpage

\end{document}